%% file: main.tex
\documentclass[conference]{IEEEtran}
\IEEEoverridecommandlockouts
\usepackage{cite}
\usepackage{amsmath,amssymb,amsfonts}
\usepackage{algorithmic}
\usepackage{graphicx}
\usepackage{textcomp}
\usepackage{subcaption}
\usepackage{balance}

\usepackage[dvipsnames]{xcolor}
\usepackage{todonotes,setspace}
\def\BibTeX{{\rm B\kern-.05em{\sc i\kern-.025em b}\kern-.08em
    T\kern-.1667em\lower.7ex\hbox{E}\kern-.125emX}}
\begin{document}


\newcounter{mycomment}
\newcommand{\mycomment}[2][]{
   \refstepcounter{mycomment}%
   \ifcase\numexpr \themycomment- 4*((\themycomment+3)/4 -1) \relax
   {%
      \setstretch{0.7}
      \todo[color=SeaGreen,size=\scriptsize]{%
         \textbf{[\uppercase{#1}\themycomment]:} #2}%
   }
   \or
   {%
      \setstretch{0.7}
      \todo[color=Salmon,size=\scriptsize]{%
         \textbf{[\uppercase{#1}\themycomment]:} #2}%
   }
   \or
   {%
      \setstretch{0.7}
      \todo[color=Thistle,size=\scriptsize]{%
         \textbf{[\uppercase{#1}\themycomment]:} #2}%
   }
   \else
   {%
      \setstretch{0.7}
      \todo[color=Tan,size=\scriptsize]{%
         \textbf{[\uppercase{#1}\themycomment]:} #2}%
   }
   \fi
}

\title{Probabilistic Neural Circuits leveraging AI-Enhanced Codesign for Random Number Generation
}

 \author{\IEEEauthorblockN{Suma G. Cardwell\IEEEauthorrefmark{1},
 		Catherine D. Schuman\IEEEauthorrefmark{3}
 		J. Darby Smith\IEEEauthorrefmark{1}, 
 		Karan Patel\IEEEauthorrefmark{3},
 		Jaesuk Kwon \IEEEauthorrefmark{2},
 		Samuel Liu \IEEEauthorrefmark{2},\\
       Christopher Allemang\IEEEauthorrefmark{1}, 
 		Shashank Misra\IEEEauthorrefmark{1},
 		Jean Anne Incorvia \IEEEauthorrefmark{2} and
 		James B. Aimone\IEEEauthorrefmark{1}} \\
 	\IEEEauthorblockA{\IEEEauthorrefmark{1}\textit{Sandia National Laboratories, Albuquerque, NM, USA} \\
 		\IEEEauthorrefmark{2}\textit{University of Texas, Austin}\\
 		\IEEEauthorrefmark{3}\textit{University of Tennessee, Knoxville}\\
 			\\
 		Email: sgcardw@sandia.gov, cschuman@utk.edu, jbaimon@sandia.gov}
		
 	}


\maketitle

\begin{abstract}
\input{abstract}

\end{abstract}

\begin{IEEEkeywords}
Probabilistic Devices, Probabilistic Neural Circuits, AI-Enhanced Codesign, Codesign, Neuromorphic computing
\end{IEEEkeywords}

\begin{figure}[ht]
	\centering
	\includegraphics[width=0.45\textwidth]{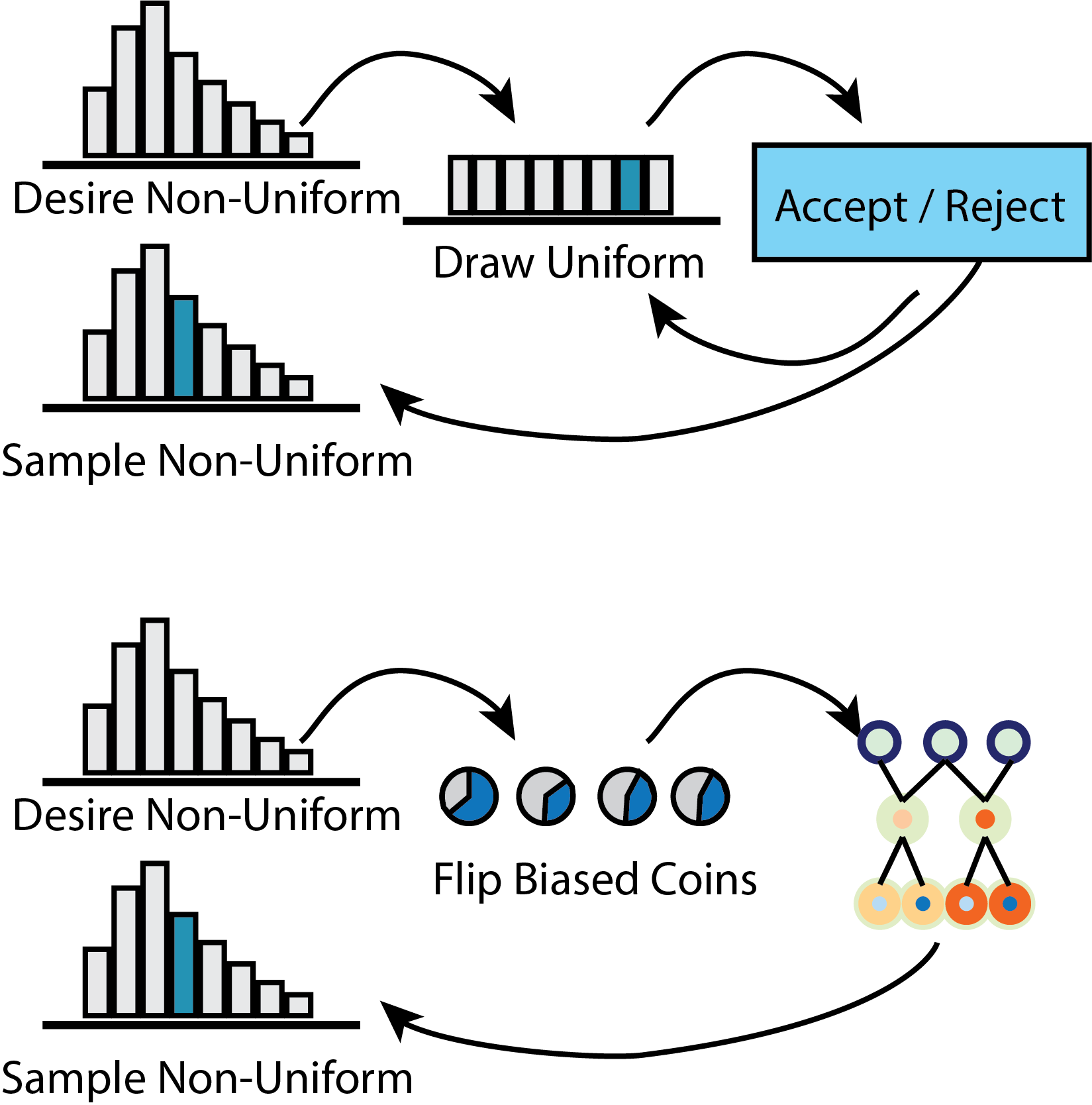}
	
	\caption{Top: In scientific computing applications today, uniform random numbers need to be converted to distributions of interest through expensive rejection sampling or related techniques. Bottom: By using device stochasticity and co-designed circuits, we can directly convert biased coin flips to our distributions of interest, avoiding repeated sampling loops.}
	\label{codesign}
\end{figure}

\section{Introduction}
Neuromorphic computing is an emerging paradigm that promises to alleviate the challenges faced by current classical computing approaches by emulating key computational principles from the brain. Recent work has shown that neural networks can be used to sample probabilistic graphs \cite{huang2014neurons, buesing2011neural}. However, today’s neuromorphic platforms have at best modest pseudo-random number generator (PRNG) capabilities. In contrast, every synapse in the brain exhibits stochastic vesicle release. Neuroscientists have observed that stochasticity at the synapse and circuit scales allows for both synaptic development and circuit functional dynamics, phenomena that are crucial for higher-level cognitive functions. Recent theoretical work has also demonstrated that neural algorithms can leverage RNGs in parallel to provide added capabilities to probabilistic algorithms while leveraging the energy and time advantages of neuromorphic parallelism \cite{smith2022neuromorphic}.
We seek to leverage stochasticity in computing by exploiting the underlying physics of emerging random number generator (RNG) devices to build probabilistic neural architectures. We will leverage stochasticity in computing, by making stochasticity ubiquitous and, crucially, making it useful. 

In many scientific computing applications today, uniform pseudo-random numbers are converted to a desired distribution. When examining a particular collider physics simulation \cite{pierog2015epos}, on the order of 270K uniform pseudo-random numbers are used to simulate a single event, and up to billions of events must be simulated to understand experimental data \cite{misra2022}. The CPU time for PRNGs in such simulations is on the order of $40-50 \%$ of the total compute time \cite{misra2022}. Contributing and adding to this cost is the necessity to convert uniform samples to distributions of interest: as PRNGs typically produce uniformly distributed numbers, it is necessary to convert these samples into distributions of interest using techniques such as rejection sampling or Markov Chain Monte Carlo algorithms (Fig. \ref{codesign}, top). 

Direct random number generation leveraging stochastic devices can promise significant energy savings for such applications. 
In this work, we explore drawing samples directly from a simple, non-uniform distribution using true random number generating (TRNG) device models (Fig. \ref{codesign}, bottom). We choose a simple distribution among four outcomes that cannot be represented by the outcome of two independent coin tosses alone. Rather, we employ a probabilistic mixing through a ``Hidden Dependence'' process that can be represented as a probabilistic circuit. The hidden process randomly selects between various coin distributions. Though our choice of distribution and hidden process circuit may seem arbitrary, they are important as they demonstrate a methodology to correlate output among independent tosses. Leveraging correlation of output is critical in direct drawing from non-trivial distributions.
This work showcases our 
results from developing neural probabilistic circuits using AI-enhanced codesign leveraging emerging neuromorphic devices for random number generation from a given distribution as illustrated in Fig. \ref{codesign}.

\section{AI-enhanced Codesign}
\label{sec:ai-enhanced-codesign}
\input{eons}

\section{Devices}
\label{sec:devices}
We evaluated two devices for random number generation, namely Tunnel Diodes (TD) and Magnetic Tunnel Junctions (MTJs). We included functional models with abstracted device characteristics to evaluate the probabilistic circuit developed by the AI-enhanced codesign framework.
\input{devices}


\section{Probabilistic Circuits}
\label{sec:prob_circuits}
\input{prob_circuits}

\section{AI-enhanced Probabilistic Circuits}
\label{sec:ai_probs_circuit}
\input{ai_prob_circuit}

\section{Future Work}
\label{sec:future_work}
In this work, we leveraged LEAP for parameter optimization; however, LEAP does not allow for network topology optimization.  For future work, for both topology and parameter optimization, we will leverage Evolutionary Optimization for Neuromorphic Systems (EONS)~\cite{schuman2016evolutionary, schuman2020evolutionary} for evolving both non-neuromorphic and neuromorphic circuits of these devices. In this work, we specifically explored a ``Hidden Dependence'' method. Future work will include other methods to evaluate correlation and rank them based on performance tradeoff. We will also explore the tradeoffs in circuit size, number of devices, sampling rates, topology, and performance. We also plan to expand our optimization technique to include system design and architecture design. 

Modeling complex problems such as nuclear and high energy physics, climate models with high precision, and novel AI techniques require simulating probabilistic behaviors. However, doing so in conventional digital hardware has a high energy cost. In this work, we explored drawing samples directly from a simple, non-uniform distribution using true random number generating (TRNG) device models. We chose a simple distribution among four outcomes that cannot be represented by the outcome of two independent coin tosses alone. Rather, we employed probabilistic mixing through a hidden process that can be represented as a probabilistic circuit.  Our choice of distribution was important to demonstrate a methodology to correlate output among independent tosses. Leveraging correlation of output is critical in direct drawing from non-trivial distributions. We demonstrated this method using three device models, namely MTJ-SHE, MTJ-VCMA and TD. Direct random number generation leveraging nanoscale devices promise significant energy and latency advantages for scientific computing applications.
    
    Our ultimate goal is to achieve a billion random number per microsecond using TRNG devices and computing on distributions with low latency and energy footprint as seen in the brain \cite{misra2022}. The results presented in this paper are a step in that direction. Our research intends to develop probabilistic neural computing, which will be relevant for both AI and scientific computing applications. 

\section*{Acknowledgment}


The authors acknowledge financial support from the DOE Office of Science (ASCR / BES) for our Microelectronics Co-Design project COINLFIPS. Sandia National Laboratories is a multimission laboratory managed and operated by National Technology \& Engineering Solutions of Sandia, LLC, a wholly owned subsidiary of Honeywell International Inc., for the U.S. Department of Energy’s National Nuclear Security Administration under contract DE-NA0003525.

This article has been authored by an employee of National Technology \& Engineering Solutions of Sandia, LLC under Contract No. DE-NA0003525 with the U.S. Department of Energy (DOE). The employee owns all right, title and interest in and to the article and is solely responsible for its contents. The United States Government retains and the publisher, by accepting the article for publication, acknowledges that the United States Government retains a non-exclusive, paid-up, irrevocable, world-wide license to publish or reproduce the published form of this article or allow others to do so, for United States Government purposes. The DOE will provide public access to these results of federally sponsored research in accordance with the DOE Public Access Plan https://www.energy.gov/downloads/doe-public-access-plan.

This paper describes objective technical results and analysis. Any subjective views or opinions that might be expressed in the paper do not necessarily represent the views of the U.S. Department of Energy or the United States Government. SAND Number: SAND2022-16607 C.

\bibliographystyle{IEEEtran}
\balance
\bibliography{references}

\end{document}

%% file: abstract.tex
Stochasticity is ubiquitous in the world around us. However, our predominant computing paradigm is deterministic. Random number generation (RNG) can be a computationally inefficient operation in this system especially for larger workloads. 
Our work leverages the underlying physics of emerging devices to develop probabilistic neural circuits for RNGs from a given distribution. However, codesign for novel circuits and systems that leverage inherent device stochasticity is a hard problem. This is mostly due to the large design space and complexity of doing so. It requires concurrent input from
multiple areas in the design stack from algorithms, architectures,
circuits, to devices. In this paper, we present examples of optimal
circuits developed leveraging AI-enhanced codesign techniques
using constraints from emerging devices and algorithms. Our
AI-enhanced codesign approach accelerated design and enabled interactions between
experts from different areas of the microelectronics design stack
including theory, algorithms, circuits, and devices. We demonstrate optimal probabilistic neural circuits using magnetic tunnel junction and tunnel diode devices that generate an RNG from a given distribution.

%% file: eons.tex

It is not immediately clear how to design circuits and tune device parameters with emerging devices.  To enable the design of circuit topology and circuit parameters, we use \textbf{\textit{evolutionary optimization}}.  There are several key reasons that we leverage evolutionary optimization for this task.  First, it can be used to optimize multiple objectives; in this case, we may be interested in how accurate the circuit performs the task, while simultaneously minimizing latency and/or energy usage. Second, it can be used to evolve both the topology of the circuit and the parameters simultaneously.  In this work, we leverage the Library of Evolutionary Algorithms in Python (LEAP)~\cite{coletti2020library} to evolve the parameters of a probabilistic circuit, though in future work, for both topology and parameter optimization, we will leverage Evolutionary Optimization for Neuromorphic Systems (EONS)~\cite{schuman2016evolutionary, schuman2020evolutionary} for evolving both non-neuromorphic and neuromorphic circuits of these devices. Third, evolutionary algorithms can potentially be used to creatively discover new ways to design circuits or leverage the underlying device characteristics.  Evolutionary algorithms have been used for decades to design analog circuits~\cite{mattiussi2007analog, zebulum1998analog}. They have been found to have surprising creativity in the design of novel solutions to a variety of problems~\cite{lehman2020surprising}.  In this work, we use LEAP's traditional evolutionary algorithm with a real representation, where the parameters to be evolved represent parameters associated with the devices themselves (described in detail in Section \ref{sec:devices}).  

%% file: devices.tex
\subsection{Tunnel Diodes}
\input{TD}

\subsection{Magnetic Tunnel Junction}
\input{MTJ}

%% file: TD.tex
\begin{figure}[h!]
	\centering
	\includegraphics[width=0.5\textwidth]{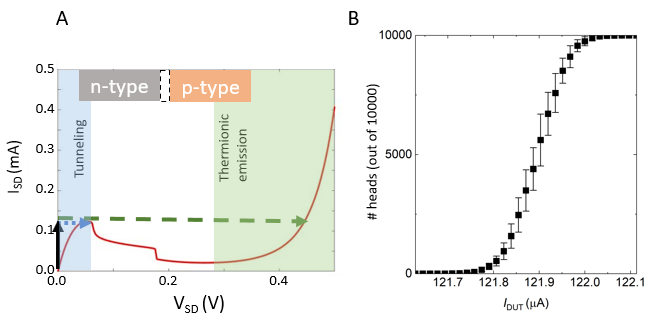}
	
	\caption{Tunnel Diode characteristics. (a) Schematic diagram of a tunnel diode, and transport curve taken from a discrete component tunnel diode (MP1103 from M-Pulse Microwave). The regions of tunneling current and thermionic emission are shown. (b) Pulsing the current near the peak (black arrow in (a)) produces the stated probability of being found in either the low voltage or high voltage branch. Error bars are associated with the number of samples.}
	\label{fig:tunnel_diode}
\end{figure}

The venerable tunnel diode (TD) has historically been used in high-speed analog applications, and is a great candidate for a practical nanoscale random number generator. They can be integrated into a standard CMOS (Complementary Metal-Oxide-Semiconductor) fabrication process, can be shrunk to nanoscale dimensions \cite{Schmid2012}, and have CMOS-compatible current densities and voltages \cite{chung2006RITD}. As shown in Figure ~\ref{fig:tunnel_diode} (a), a TD consists of a strongly n-doped and p-doped junction, and conducts either by tunneling through or by thermionic emission over the narrow depletion region. Under current bias, the random occupancy of charge traps in the depletion region determine whether the device conducts through tunneling ($\sim$ 0.05 V) or thermionic emission ($\sim$ 0.45 V). Interpreting tunneling as tails, and thermionic emission as heads, tuning the TD can be accomplished by tweaking the bias current near the region of the peak, where lower currents bias the TD towards tunneling, and higher currents towards thermionic emission, as shown in Figure ~\ref{fig:tunnel_diode} (b). The resulting coin flips have been shown to pass established tests of statistical quality \cite{bernardo2017extracting}. While the data shown in Figure ~\ref{fig:tunnel_diode} was taken with a large discrete component device at the MHz speed of our data acquisition, both scaled and high-speed devices have been demonstrated in the literature. The device in Ref. ~\cite{Schmid2012} has a dimension under 100 nm, and consumes a power of 13 nW for tunneling, and 38 nW for thermionic emission. The time to generate a flip can be determined by the time taken to charge the junction capacitance (0.25 fF) twice - once to reset it, and once to charge it for the coin flip - 1.2 ns. Assuming no additional overhead from parasitics, this produces an energy consumption of 0.05 fW for a heads, and 0.02 fW for a tails. The device in Ref. ~\cite{chung2006RITD} worked at time scales as short as 50 ps, but is larger at 600 nm a side. The correspondingly larger 0.8 mA current makes it simpler to integrate, with the short operation time counteracting the larger power dissipation, limiting the energy consumption to 50 fJ for heads and 20 fJ for tails. We use this last TD device for resource estimation later in this manuscript.

%% file: MTJ.tex
\begin{figure}[b!]
	\centering
	\includegraphics[width=0.5\textwidth]{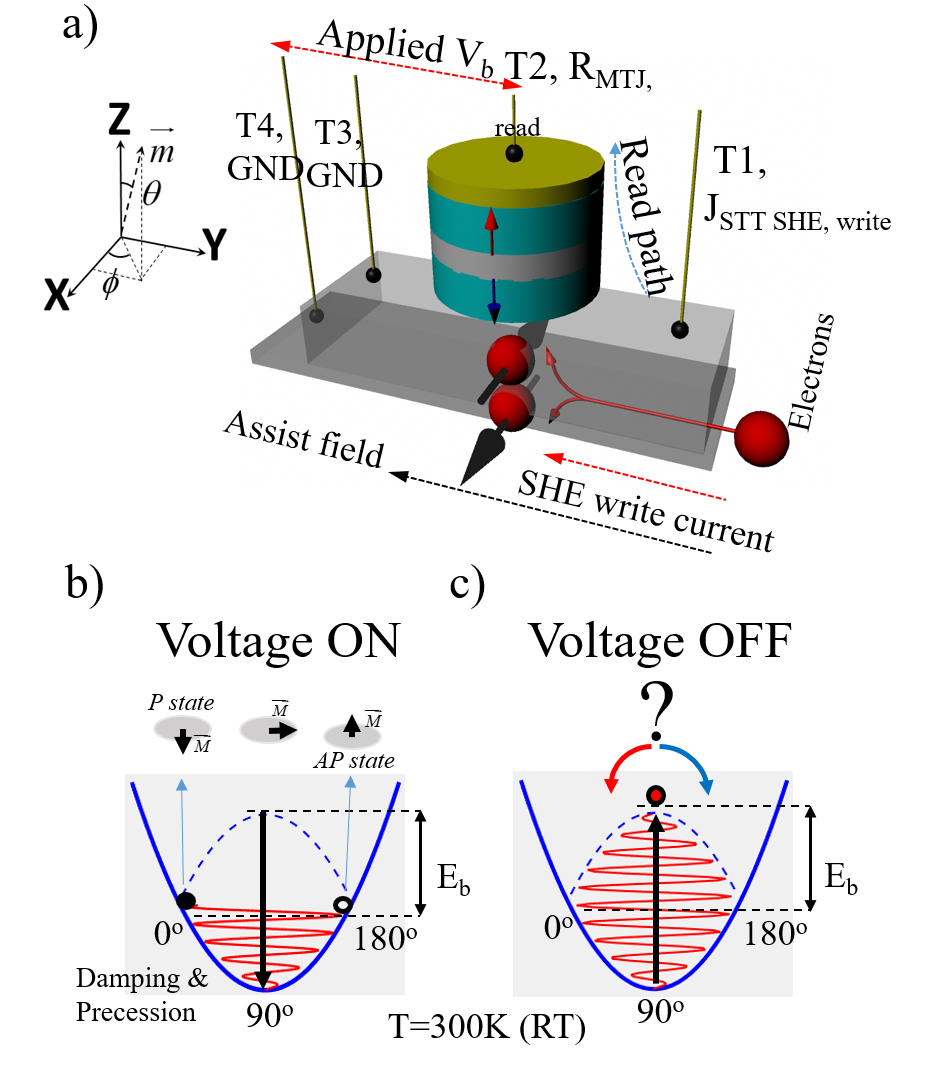}
	
	\caption{a) Schematic illustration of the VCMA-MTJ stochastic device structure with various knobs to control the switching probability. T2-T4 applies a biased voltage on the MTJ structure. Spin transfer torque (STT) (T1-T2) or spin Hall effect (SHE) (T1-T3) can bias the switching probability. b) A and AP states are stable without applying a voltage. A biased voltage eliminates stable states with a precessional switching to the middle in-plane state. c) After removing the bias voltage, the device recovers to the stable states with an unpredicted switching towards A or AP state.}
	\label{MTJfig}
\end{figure}

Magnetic tunnel junctions (MTJs) have recently become a crucial component of memory, in-memory computing, and probability bit (p-bit) device applications \cite{camsari2019,camsariNature,hayakawa2021}. An MTJ consists of an insulating tunnel barrier between two thin ferromagnetic layers; due to spin-dependent electron transport, the MTJ has a high and low resistance state depending on the P (parallel) and AP (anti-parallel) orientation of the magnetization of the two ferromagnetic layers. One of these layers is held at a fixed magnetization, while the other, the free layer, can be switched via voltage, current, and heat. The thermally-driven, stochastic nature of the switching of the free layer can lead to generation of streams of random bits \cite{Safransk2021}. The MTJ can be set up as a stochastic read device, by having a low-anisotropy magnetic free layer, near the superparamagnetic limit, that randomly switches its magnetization at room temperature. The random fluctuations of the MTJ resistance can then be read. Alternatively, the MTJ can be set up as a stochastic write device, with a stable-anisotropy free magnetic layer that has a probability of switching its magnetization, and therefore the MTJ resistance, depending on the amplitude and duration of an applied current pulse. In both of these configurations, the weight of the probabilistic bit, i.e. how much time it spends in P or AP states, can be controlled using constant applied fields and/or DC bias voltages or currents. As an alternative to spin transfer torque (STT), the spin Hall effect (SHE) can be used to change the state of the MTJ with low-energy switching behavior as shown in Fig. \ref{MTJfig} (a). We will refer to these devices as MTJ-SHE.

An alternative way to generate stochastic bit streams with MTJs is using voltage-controlled magnetic anisotropy (VCMA), where the anisotropy of the free magnetic switching layer of the MTJ is modulated using voltage; this effect has been extensively explored for memory applications. The VCMA effect can exert unpredictable magnetization dynamics on the free layer at room temperature, but it has not yet been fully explored for generating controllable random bit streams. Here, we build numerical models of both the VCMA-MTJs and the MTJ-SHE based on the Landau-Lifshitz-Gilbert (LLG) equation \cite{Wang2015,Leliaert2017}, modeling a standard perpendicular MTJ stack comprised of CoFeB (free layer)/MgO/CoFeB (fixed layer), as shown in Fig. \ref{MTJfig}a. Fig. \ref{MTJfig}b-c depicts the operation of the device using VCMA to generate a random bit stream: first, a voltage is applied that reduces the perpendicular magnetic anisotropy and sends the magnetization from out of plane to in-plane through precessional oscillations (Fig. \ref{MTJfig}b). Then, the voltage is turned off, and when the anisotropy pops back, the magnetization must choose one of two out-of-plane directions to stabilize the MTJ in either a P or AP state (Fig. \ref{MTJfig}c). We will refer to this device as MTJ-VCMA. Similarly, the MTJ-SHE uses the SHE to send the magnetization to a middle-state when the SHE is on, and subsequently the magnetization must recover to either a P or AP state.

The developed abstract models for both the MTJ-SHE and MTJ-VCMA capture the devices functionality and energy consumption. These device models are utilized when evaluating the fitness function of our evolutionary method and generated bit stream from this analytical code is provided to the probabilistic circuit.

%% file: prob_circuits.tex
An 
important application for probabilistic circuits is to generate RNGs from a given distribution. There are many potential methods for taking a series of two-state random outcomes, such as those produced by an MTJ or TD, and converting them into a sample from a desired distribution. One approach would be to utilize classic flipping methods \cite{gryszka2021biased}. Though, taking such approaches may lead to tossing an unbounded number of arbitrarily tuned coins \cite{gryszka2021biased}.  Another approach could come from identifying outcomes of a series of tossed to coins to either a discrete distribution or a discretized continuous distribution. Such an approach would guarantee a finite number of tosses provided the distributions have finite support or are truncated in some fashion. We will focus on sampling from discrete, finitely supported distributions using such an identification. We will not review all such methods of sampling nor argue that this is a  superior one in any metric. However, this 
test case demonstrates an important feature: we are able to produce correlation among independent coin tosses to sample from a distribution that cannot be sampled from two independent coin flips alone.

To mimic coin tossing, we consider the Bernoulli distribution. This distribution is defined on two events, $0$(T) and $1$(H). We will say $\mathbb{P}(1)=p$ and $\mathbb{P}(0)=1-p$. If we wish to sample from a distribution with just two events, or a discretized continuous distribution with two effective events, we need only make an identification of $H$ with event 1 and $T$ with event 2 and assign the probability $p$ accordingly.  Note, for a discretized continuous distribution, this would mean assigning $p$ to be the integral of the probability density function (pdf) over the bin corresponding to event 1.   If we have more than two events, more care may need to be taken.  For example, if we have four events, we may toss two coins in a row.  Let $p$ be the probability of heads for coin 1 and $q$ be the probability of heads for coin 2. Since there are four distinct outcomes (HH, HT, TH, and TT), we need only solve the appropriate system of equations to solve for the probabilities $p$ and $q$.  If you have three events, however, you will need to arbitrarily select two of the four outcomes to represent a single one. In such instances there may be many non-unique representations. This methodology extends easily to any finite number of states.

However, situations may arise where there are no solutions for the coin heads probabilities. In these situations, we will need to enforce correlation among the outcomes of the coin tosses. 
There are many ways to correlate coin flips. One way to correlate the output of coin flips
among Bernoulli random variables is through a hidden dependence model.

The ``Hidden Dependence'' model creates dependent random variables that are connected by some underlying, possibly hidden, random variable as illustrated in Fig. \ref{fig:hidden_correlation}.  This hidden variable can be thought of in many ways.  It could, for instance, represent some physical model. The dependence may induce correlation among outputs.  Relating this to COINFLIPS, the hidden variable could be the physics of a device such as an MTJ or a TD. In our test case, the hidden variable could be another coin whose outcome determines which set of two additional coins are flipped. Hidden dependence coins can be flipped simultaneously or in any order. This means from an outside viewpoint, the coins appear to be independent tosses even though they represent dependent random variables.

\subsection*{Hidden Dependence Bernoulli Coins}
\begin{figure}[ht]
	\centering
	\includegraphics[width=0.48\textwidth]{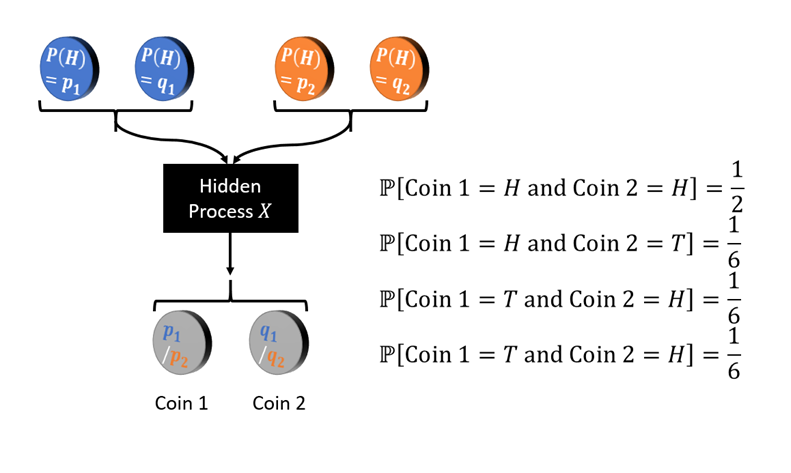}
	\caption{In the ``Hidden Dependence'' model, there is a hidden process, stochastic or deterministic, that controls the probability of heads among a collection of coins. In this cartoon, the hidden process chooses which set of coins is flipped. The observer only sees a single set, the effective flipping set. 
    }
	\label{fig:hidden_correlation}
\end{figure}

 In the ``Hidden Dependence'' scenario, there is a hidden process, stochastic or deterministic, that controls the probability of heads among a collection of coins, as shown in Fig. \ref{fig:hidden_correlation}.

The hidden process that controls the dependence among the effective visible Bernoulli coins could be one of many options. It could be a deterministic process that shuffles the used coin set; it could be a knob that tunes the probability of heads among all coins; or it could be some more complicated random determination. However, the key feature for this scenario is that the effective visible set of coins can be flipped in any desired order (or all at once). Any causation of output correlation and dependence is absorbed by something a viewer does not see.

As an analogy, one could imagine a person that has two sets of coins in their pocket, each set has a coin labeled coin 1 and a coin labeled coin 2.  The person encounters a series of individuals.  When encountering an individual, the person selects a set of coins and the individual is obliged to flip them.  If the person is strategic in how they choose the set of coins for each individual to flip, they can induce a correlation in the effective output of coin 1 and coin 2.


\begin{figure*}[t!]
\centering
\begin{subfigure}{.32\textwidth}
  \centering
  \includegraphics[width=1.0\linewidth]{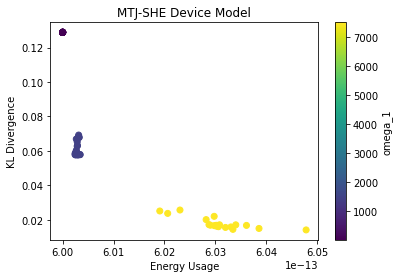}
  \caption{$\omega_1$}
  \label{fig:sfig1}
\end{subfigure}
\begin{subfigure}{.32\textwidth}
  \centering
  \includegraphics[width=1.0\linewidth]{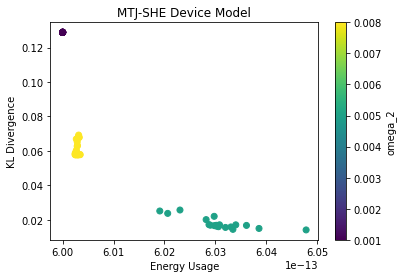}
  \caption{$\omega_2$}
  \label{fig:sfig2}
\end{subfigure}
\begin{subfigure}{.32\textwidth}
  \centering
  \includegraphics[width=1.0\linewidth]{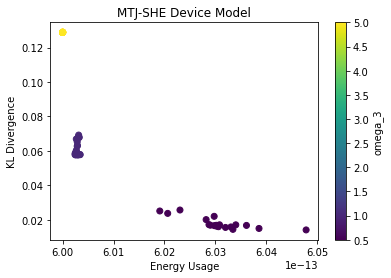}
  \caption{$\omega_3$}
  \label{fig:sfig3}
\end{subfigure}
\caption{Impact of multi-objective weights $\omega_1$, $\omega_2$, $\omega_3$ on the KL divergence and energy usage of MTJ-SHE devices.}
\label{fig:moo_weight_effect}
\end{figure*}

We now define a concrete example that we will develop a circuit for.  Suppose the distribution we would like to sample from is that of a four-sided die that rolls $0$ with probability $1/2$, and rolls 1, 2, 3 with probability 1/6 each.  It's easy enough to identify these outcomes with the outcomes of two coin flips: 0 corresponds to HH, 1 to HT, 2 to TH, and 3 to TT.  If $p$ is the probability of $H$ for coin 1 and $q$ is the probability of H for coin 2, then we need the following to hold:
\begin{align}
\begin{split}
pq &=\displaystyle\frac{1}{2},\\
p(1-q) &=\displaystyle\frac{1}{6},\\
(1-p)q &=\displaystyle\frac{1}{6},\\
(1-p)(1-q)&=\displaystyle\frac{1}{6}.
\end{split}
\label{eq:bad_dist}
\end{align}
Despite the simple nature of this distribution, it is a quintessential example of a system that requires the use of a hidden process. As hinted above, there is a problem with this setup. This system of equations has no solution for $p$ or $q$. In order to realize the desired die roll, we will need to employ a mechanism that alters the probabilities $p$ and $q$.

The set of equations \eqref{eq:bad_dist} are ill defined as $p$ and $q$ are over-constrained. To rectify this, we can relax the constraints by adding in additional variables, namely by adding in a second set of coins. Returning to our heuristic, we want our person handing out the sets of coins to hand out set 1 for a fraction $w$ of the time and set 2 for the remainder of the time. Set 1 has coin 1 with probability of heads $p_1$ and has coin 2 with probability of heads $q_1$. Set 2 has probabilities of heads $p_2$ and $q_2$ respectively. This yields the following system of equations.

\begin{align}
\begin{split}
wp_1q_1 +(1-w)p_2q_2 &= \displaystyle\frac{1}{2},\\
wp_1(1-q_1)+(1-w)p_2(1-q_2)&=\displaystyle\frac{1}{6},\\
w(1-p_1)q_1+(1-w)(1-p_2)q_2&=\displaystyle\frac{1}{6},\\
w(1-p_1)(1-q_1)+(1-w)(1-p_2)(1-q_2)&=\displaystyle\frac{1}{6}
\end{split}
\label{eq:rectified_set}
\end{align}

By altering between coin sets with probability $w$, we have induced a new hidden process that controls the observed outcomes among the effective observed coins. Notably, this introduction allows the system to be solved for some choices of $w$.  However, given that we have a choice in $w$, this means we may not have a unique global solution for $p_1$, $p_2$, $q_1$, and $q_2$. 

While we concede that this exemplar is not tied to any specific application, any practical distribution will not likely have outcome probabilities that may be represented in a system similar to \eqref{eq:bad_dist}. This exemplar is meant to showcase how one might correlate output and generate circuits to deal with such problems. If the future of probabilistic computer design can use distributions formed in such a manner, we will need the algorithmic expertise to handle these types of over-constrained situations.

We model equation set \eqref{eq:rectified_set} in Section \ref{sec:ai_probs_circuit}, and include the functional models of the various emerging devices we studied in this work.

%% file: ai_prob_circuit.tex
We modeled the equations described in Section \ref{sec:prob_circuits} and leveraged LEAP to generate the hidden weight ($w$) and probability weights ($p_1$, $q_1$, $p_2$, and $q_2$) for the coinflip devices, where $w, p_1,q_1,p_2,q_2 \in [0,1]$. For our examples we used LEAP's traditional evolutionary algorithm with a real representation, where the parameters to be evolved represent parameters associated with emerging devices, namely TD, MTJ-SHE, and MTJ-VCMA devices.  We used tournament selection, uniform crossover, and Gaussian mutation with a standard deviation of $0.001$ and one expected mutation per individual. For our use-case this is to find the optimal value of $w$, $p_1$,$q_1$, $p_2$, and $q_2$ such that our desired objective is achieved.  In this case, we have three key objectives, described in our fitness function below. 

	

	

\textbf{Fitness function}: The fitness function defines the objective of the evolutionary algorithm.  In this work, we leverage a multi-objective fitness function, where we seek to have LEAP find the appropriate values for these that result in minimizing the KL divergence value from the desired distribution while simultaneously minimizing the difference between the weights of coins and 0.5 (a fair coin), as well as minimizing energy usage (based on the particular device). The KL divergence objective is specifically measuring the difference from the given distribution. In particular, we calculate the following:

\begin{align}
\begin{split}
v(0) &= wp_1q_1 +(1-w)p_2q_2\\
v(1) &= wp_1(1-q_1)+(1-w)p_2(1-q_2)\\
v(2) &= w(1-p_1)q_1+(1-w)(1-p_2)q_2\\
v(3) &= w(1-p_1)(1-q_1)+(1-w)(1-p_2)(1-q_2)\\
\end{split}
\label{eq:q_functions}
\end{align}

We then set $p(0) = \frac{1}{2}$, $p(1) = \frac{1}{6}$, $p(2) = \frac{1}{6}$, $p(3) = \frac{1}{6}$. \\

The KL divergence is then calculated:
\begin{equation}
    KL = \sum_{i=0}^{3} v(i)\log{\frac{v(i)}{p(i)}}
\end{equation}

For the given device model (MTJ-SHE, MTJ-VCMA or TD), we then calculate the energy usage for devices with probabilities $p_1$, $p_2$, $q_1$, and $q_2$. We sum these energy values and produce a single energy usage, $EN$.  To allow us to investigate the tradeoff between different objectives, we include three objective weights $\omega_1, \omega_2, \omega_3$.  Thus, our overall fitness function is:

\begin{equation}
\begin{split}
    f(w, p_1, p_2, q_1, q_2) & =  \omega_1 KL(w, p_1, p_2, q_1, q_2) \\
    & + \omega_2 \left ( \sum_{i=1}^{2}|p_i-0.5|+\sum_{i=1}^2 |q_i-0.5| \right ) \\
    & + \omega_3 EN(p_1, p_2, q_1, q_2)
    \end{split}
\end{equation}

\subsection{Tradeoffs and Performance Evaluation}

In Fig.~\ref{fig:moo_weight_effect}, we can see the impact on the KL-divergence and the energy use for different values for the three objective weights $\omega_1$, $\omega_2$, $\omega_3$ for the MTJ-SHE device.  Here, we can clearly see that by setting the values for each weight differently, we can achieve different varying levels of KL-divergence and energy usage.  For example, a larger $\omega_1$ for the minimizing KL-divergence objective and a smaller $\omega_3$ for the minimizing energy usage objective result in sets of device parameters that achieve smaller KL-divergence values at the expense of using more energy. For future co-design efforts, we can leverage this multi-objective approach to tune the performance of our devices to achieve our desired tradeoffs in performance and energy usage. We omit the plots for MTJ-VCMA and TD, where there is less impact of the weight of the coins on the energy usage. In the remainder of the paper, we focus on primarily minimizing KL divergence, resulting in a focus on the following weight sets: $\omega_1=7500$, $\omega_2= 0.005$, and $\omega_3=0.5$.

\subsection{MTJ-VCMA Results}

\begin{figure}[ht!]
    \centering
    \includegraphics[width=0.45\textwidth]{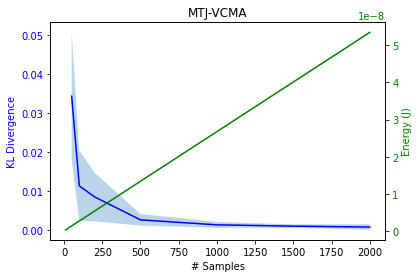}
    \caption{KL divergence and energy usage vs. number of samples for the given distribution with the MTJ-VCMA device. There were 10 trials conducted per sample.  Sample sizes investigated include 10, 50, 100, 200, 500, 1000, 1500 and 2000.}
    \label{fig:mtj_vcma_samples_kl_energy}
\end{figure}

\begin{figure}[ht!]
    \centering
    \includegraphics[width=0.47\textwidth]{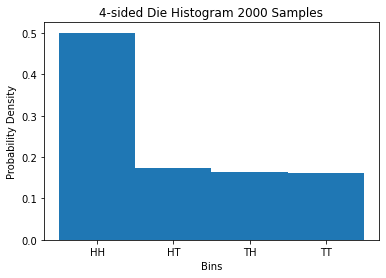}
    \caption{Empirical distribution using MTJ-VCMA for 2000 samples in a single run. The optimized values generated were: $w=0.306$, $p_1=0.448$, $p_2=0.746$, $q_1=0.439$, and $q_2=0.749$.}
    \label{fig:mtj_vcma_distribution}
\end{figure}

We selected the device configuration for MTJ-VCMA that gave the lowest KL-divergence value. This device was optimized using $\omega_1=7500$, $\omega_2= 0.005$, and $\omega_3=0.5$.  The resulting parameters found through optimization over 1000 generations in LEAP for this device were $w=0.233$, $p_1=0.237$, $p_2=0.781$, $q_1=0.199$, and $q_2=0.805$. 
Fig.~\ref{fig:mtj_vcma_samples_kl_energy} plots the KL divergence and energy usage for the $n$ samples for this MTJ-VCMA device across 10 different sets of samples. As seen Fig. \ref{fig:mtj_vcma_samples_kl_energy}, increasing the number of samples lowers the KL-divergence from the desired probability distribution. However, more samples come at the cost of increased energy consumption.  One of the resulting distributions with 2000 samples is shown in Fig.~\ref{fig:mtj_vcma_distribution}. 

\subsection{MTJ-SHE Results}

\begin{figure}[t!]
    \centering
    \includegraphics[width=0.45\textwidth]{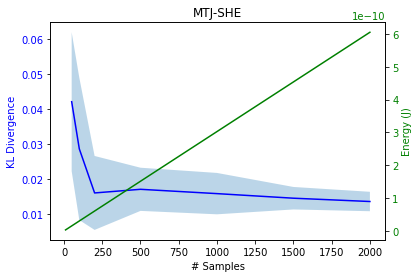}
    \caption{KL divergence and energy usage vs. number of samples for the given distribution with the MTJ-SHE device. There were 10 trials conducted per sample.  Sample sizes investigated include 10, 50, 100, 200, 500, 1000, 1500 and 2000.}
    \label{fig:mtj_she_samples_kl_energy}
\end{figure}

\begin{figure}[h!]
    \centering
    \includegraphics[width=0.45\textwidth]{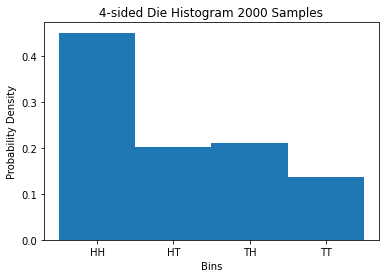}
    \caption{Empirical distribution using MTJ-SHE for 2000 samples in a single run. The optimized values generated were: $w=0.306$, $p_1=0.448$, $p_2=0.746$, $q_1=0.439$, and $q_2=0.749$.}
    \label{fig:mtj_she_distribution}
\end{figure}

We also selected the device configuration for MTJ-SHE that gave us the lowest KL-divergence value.  This device was optimized using $\omega_1=7500$, $\omega_2= 0.005$, and $\omega_3=0.5$.  The resulting parameters found through optimization over 1000 generations in LEAP for this device were $w=0.306$, $p_1=0.448$, $p_2=0.746$, $q_1=0.439$, and $q_2=0.749$. 
Fig.~\ref{fig:mtj_she_samples_kl_energy} plots the KL divergence and energy usage for the $n$ samples for this MTJ-SHE device across 10 different sets of samples. As seen Fig. \ref{fig:mtj_she_samples_kl_energy}, increasing the number of samples lowers the KL-divergence from the desired probability distribution. However, more samples come at the cost of increased energy consumption.  One of the resulting distributions with 2000 samples is shown in Fig.~\ref{fig:mtj_she_distribution}. 

It is observed that there is an orders of magnitude energy reduction when SHE is used compared to VCMA. In practice, SHE can require a DC external magnetic field. In the model used, the SHE current is polarized perfectly in-plane; experimentally, this is not always the case. The VCMA effect could provide robustness at the cost of energy, since when VCMA is used, the magnetization is dependent on anisotropy instead of spin current.

\subsection{TD Results}

\begin{figure}[h!]
    \centering
    \includegraphics[width=0.45\textwidth]{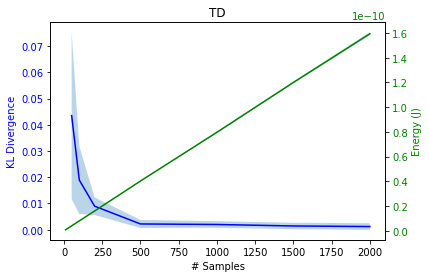}
    \caption{KL divergence and energy usage vs. number of samples for the given distribution with the TD device. There were 10 trials conducted per sample.  Sample sizes investigated include 10, 50, 100, 200, 500, 1000, 1500 and 2000.}
    \label{fig:td_samples_kl_energy}
\end{figure}

\begin{figure}[h!]
    \centering
    \includegraphics[width=0.46\textwidth]{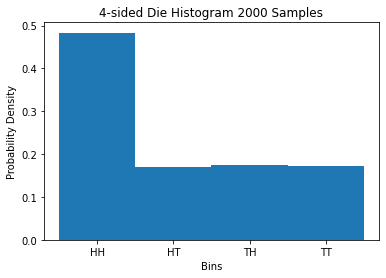}
    \caption{Empirical distribution using TD for 2000 samples in a single run. The optimized values generated were: $w=0.714$, $p_1=0.891$, $p_2=0.107$, $q_1=0.766$, and $q_2=0.419$.}
    \label{fig:td_distribution}
\end{figure}

Again, we selected the device configuration for TD that gave us the lowest KL-divergence value.  This device was optimized using $\omega_1=7500$, $\omega_2= 0.005$, and $\omega_3=0.5$.  The resulting parameters found through optimization over 1000 generations in LEAP for this device were $w=0.714$, $p_1=0.891$, $p_2=0.107$, $q_1=0.766$, and $q_2=0.419$. Fig.~\ref{fig:td_samples_kl_energy} plots the KL divergence and energy usage for the $n$ samples for this TD device across 10 different sets of samples. As seen Fig. \ref{fig:td_samples_kl_energy}, increasing the number of samples lowers the KL-divergence from the desired probability distribution. However, more samples come at the cost of increased energy consumption.  One of the resulting distributions with 2000 samples is shown in Fig.~\ref{fig:td_distribution}. 
\begin{figure*}
    \centering
    \includegraphics[width=\textwidth]{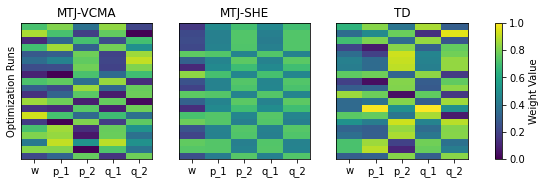}
    \caption{Optimized weight values for each device over twenty optimization runs using LEAP.}
    \label{fig:device_weight_values}
\end{figure*}

\subsection{Conclusions}
We leveraged LEAP to generate optimal device parameters for a given probability distribution given in equations \eqref{eq:rectified_set}. These parameters were then plugged into device models to generate empirical distributions as shown in Figs.~\ref{fig:mtj_vcma_distribution},  \ref{fig:mtj_she_distribution} and \ref{fig:td_distribution}.  In Fig.~\ref{fig:device_weight_values}, we show the 20 different sets of weight values that were optimized for each device type for $\omega_1=7500$, $\omega_2= 0.005$, and $\omega_3=0.5$.  Here, we see that the weights are customized for the device's behavior to target the best performance in terms of KL divergence and energy usage.


	

One of the challenges in optimizing for both algorithms and devices was appropriately abstracting the device models and algorithmic constraints. The functional models developed will also be evolved in time as new device data and research emerges. Our framework is set up so as to accommodate such changes. 